\documentclass[a4paper,fleqn,usenatbib]{mnras}

\usepackage[T1]{fontenc}
\usepackage{ae,aecompl}

\usepackage{graphicx}	
\usepackage{amsmath}	
\usepackage{amssymb}	
\usepackage{xcolor}

\newcommand{\msun}{\mbox {$\mathrm M_{\odot}$}}
\newcommand{\mdot}{$\mathrm{\dot{M}}$}

\def\sna{SN~Ia}
\def\sne{SNe~Ia}

\def\apgt{\ {\raise-.5ex\hbox{$\buildrel>\over\sim$}}\ }
\def\aplt{\ {\raise-.5ex\hbox{$\buildrel<\over\sim$}}\ }
\newcommand{\pyr}{\mbox {{\rm yr$^{-1}$}}}

\title[Observational properties of \sne\ progenitors]{Observational Properties of \sne\ Progenitors Close
to the Explosion}

\author[A. Tornamb\'e et al.]{
A. Tornamb\'e,$^{1}$\thanks{E-mail: amedeo.tornambe@oa-roma.inaf.it (AT), piersanti@oa-abruzzo.inaf.it (LP),
raimondo@oa-abruzzo.inaf.it (GR), raffaele.delgrande@oa-roma.inaf.it (RD)}
L. Piersanti,$^{2,3}$
G. Raimondo,$^{2}$
R. Delgrande$^{4,5,1}$
\\
$^{1}$INAF-Osservatorio Astronomico di Roma, via Frascati, 33, I-00040, Monte Porzio Catone, Italy\\
$^{2}$INAF-Osservatorio Astronomico d'Abruzzo, via Mentore Maggini, snc, I-64100, Teramo, Italy\\
$^{3}$INFN-Sezione di Perugia, via A. Pascoli, I-06123 Perugia, Italy\\
$^{4}$Physics Department, Universit\`a di Roma ``Tor Vergata'', via della Ricerca Scientifica, 1, I-00133, Roma, Italy\\
$^{5}$ INFN-LNF, via E. Fermi 40, I-00044, Frascati, Italy
}

\date{Accepted XXX. Received YYY; in original form ZZZ}

\pubyear{2016}

\begin{document}
\label{firstpage}
\pagerange{\pageref{firstpage}--\pageref{lastpage}}
\maketitle

\begin{abstract}
We determine the expected signal in various observational bands of 
Supernovae Ia progenitors just before the explosion by assuming the 
rotating Double Degenerate scenario. 
Our results are valid also for all the evolutionary scenarios invoking 
rotation as the driving mechanism of the accretion process as well as 
the evolution up to the explosion. 
We find that the observational properties depend mainly on the mass of 
the exploding object, even if the angular momentum evolution after the 
end of the mass accretion phase and before the onset of C-burning plays 
a non-negligible role. 
Just before the explosion the magnitude $M_V$ ranges between 9 and 11 mag, while the 
colour (F225W-F555W) is about -1.64 mag. The photometric properties 
remain constant for a few decades before the explosion. During the
last few months the luminosity decreases very rapidly. 
The corresponding decline in the optical bands varies from few 
hundredths up to one magnitude, the exact value depending on both the 
WD total mass and the braking efficiency at the end of the mass 
transfer. This feature is related to the exponentially increasing 
energy production which drives the formation of a convective core 
rapidly extending over a large part of the exploding object. Also a 
drop in the angular velocity occurs. We find that observations in the 
soft X band (0.5 -2 keV) may be used to check if the \sne\ progenitors
evolution up to explosion is driven by rotation and, hence, to 
discriminate among different progenitor scenarios.
\end{abstract}




\section{Introduction}

In spite of their pivotal role in observational cosmology, type Ia
Supernovae (\sne) still remain an intriguing mystery, as no clear
consensus about their progenitor system exists so far. In fact,
the observational properties of the resulting explosive event are
  largely independent of the considered scenario for the progenitor's
  explosion, mainly because the explosion itself smears out almost
completely the previous evolutionary imprint.  It has been suggested
that some indications about the progenitors could be derived by
analyzing archival images of the site where a \sna\ explodes
\citep[e.g.][]{Li+11,Kelly+14,McCully+14}. We note 
that the imprint of the progenitor
could be found either if the donor star is still there at the epoch of
the explosion, as expected in the classical Single Degenerate (SD)
scenario \citep{whelan1973}, or some material coming from the donor
remains in the circumstellar medium, as is possible in the classical
Double Degenerate (DD) scenario \citep{iben1984}. In recent years, it
has been suggested that rotation plays a major role in determining the
evolution up to the explosion
\citep{piersanti2003a,piersanti2003b,distefano2011,hachisu2012}.  In
fact, in interacting binary systems the accretor may acquire a large 
rotational velocity owing to the continuous angular momentum deposition
via mass transfer. Rotation overcomes many shortcomings of
both the SD and DD scenarios; in addition it represents a key
ingredient in the Core Degenerate (CD) scenario by \citet{ilkov2012}. Since
rotation ``lifts'' stellar structures, it transpires that rotating
objects can increase their total mass above the canonical
Chandrasekhar mass limit and, hence, the ``rotating scenarios'' for
\sne\ progenitors allow one to explain the origin of both ``normal''
and superluminous events, the leading parameter being the total
mass of the exploding object.  An important consequence of such an
evolutionary scenario is that accreting WDs in the mass range
1.4-2.2\footnote{The upper limit is fixed by considering that in the
  DD scenario the maximum allowed mass for a CO WD is not larger than
  1.1 \msun, according to standard stellar evolution. On the other
  hand, in the SD scenario, due to the partial retention of the
  accreted matter onto the CO WD, it appears quite unrealistic that
  the mass of the exploding object could exceed such a limit
  \citep[see also][]{hachisu2012}.} \msun\, are stable as the pressure
gradient is exactly counterbalanced by the effective gravity 
(gravitational plus ``centrifugal'' forces), so that they cannot 
attain the physical conditions suitable for C-ignition in the innermost
zone. In order
to get such an outcome, it is necessary that the angular velocity
profile of the accreting WD is modified either via internal angular
momentum distribution, as occurs in differentially rotating
objects, or via angular momentum losses. Even if the exact physical
characterization of these two processes is still missing it is
reasonable to estimate that they need at least several $10^6$ yr; during
this time span all the information about the original binary systems
is lost, so that the only way to find {\it directly} an imprint of the
donor star would be to analyze astronomical frames taken at the
Pleistocene Era!

In the present work, we derive expected observational properties of SNe
Ia progenitors close to the explosion epoch by assuming that rotation
is the leading parameter driving both the mass transfer phase and the
evolution up to the Carbon ignition at the center. With this aim we
adopt the rotating DD scenario for the progenitors by
\citet{piersanti2003a,piersanti2003b} and we compute full evolutionary
models from the onset of the mass transfer process up to the dynamical
breakout, which occurs when the temperature at the burning point
approaches $\simeq 8\times 10^{8}$ K \citep{lesaffre2006}. We note 
that our analysis and findings are valid also for the rotating SD
scenario as well as for the CD scenario. In fact, when rotation is
properly taken into account, the mass transfer process does not 
directly drive  
the accreting WD to the explosion, but it continues up and
beyond the canonical Chandrasekhar mass limit until the mass reservoir
is completely exhausted. After that, the evolution up to the explosion
is determined by the angular momentum redistribution along and/or
angular momentum losses from the accreted CO WD, independently of the
previous evolution. Our analysis is not applicable to
the {\sl prompt merging} scenario by \citet{pakmor2012} as well as to
the canonical non-rotating SD scenario.

The present work is structured as follows: in \S 2 we review the
accretion phase in the rotating DD scenario, exploring the dependence
of the results on various assumptions about the physical mechanism
driving the mass accretion process; in \S 3 we discuss the physical
properties of the WD after the mass accretion has stopped and
we analyze its evolution up to the explosion; in \S 4 we derive the
expected observational properties during the last few months before the
explosion of the massive WD.  Finally, our results are summarized and
discussed in \S 5.

\section{The Mass Transfer Phase}

In the present computations we use as initial CO WD the same model with total
mass $\mathrm{M_{WD}=0.8}$ \msun\, as in \citet{piersanti2003a} and we
assume that it is embedded in a close binary system with another CO
WD with mass $\mathrm{M_{donor}=0.71}$ \msun . We compute all the
evolutionary sequences by using the F.RA.N.E.C. evolutionary code
\citet{chieffi1989}, modified according to the prescriptions in
\citet{piersanti2003a,piersanti2003b} to account for the effects of
rotation in the hypothesis of high efficiency of angular momentum
transport.

Due to gravitational wave radiation (GWR) emission the binary system shrinks and the
less massive component first overfills its own Roche lobe, giving rise
to a dynamical mass transfer so that it completely disrupts forming a
thick accretion disk around the surviving companion.  CO-rich matter
flows from the disk to the WD at a very high rate, depositing also
angular momentum. The rapid mass accretion determines 
a huge thermal energy excess and produces
a large increase of the surface angular velocity, so that the accretor
expands on a very short timescale and attains on the surface the critical angular
velocity. In this condition, no more matter can be added, so that mass
deposition stops for a while and, hence, thermal energy can be
diffused inward and the WD can contract. The accretor
recedes from the critical conditions and mass deposition can
resume. The rate at which matter could be added to
the CO WD is determined by its structural properties, in particular by
the temperature profile. Later on, when the structure becomes
thermally balanced, so that the inward heat transfer timescale becomes
longer, the effective accretion rate is determined by other
physical mechanisms. In \citet{piersanti2003b} it has been
shown that owing to the continuous angular momentum deposition, the WD
increases its rotational energy and, when the latter attains a critical
value, the accretor adopts an ellipsoidal shape, so that it acquires a
quadrupole momentum and gravitational wave radiation is 
emitted, thus braking down the WD itself.  Several different
mechanisms have been suggested by various authors to describe
the angular momentum losses from the accreting WD \citep[see e.g.][and
references therein]{andersson1995,boshkayev2014,ilkov2012}. In the present work
we do not assume any specific physical mechanism, but we 
describe such a process as an exponential law by adopting two
characteristic timescales, namely $\mathrm{\tau_B=10^{5}\ and\
  10^{6}}$ yr as representative of the braking efficiency\footnote{The
  formulation of the braking law is the same as in \citet{piersanti2003b} -
  see their Eq. 3 and the corresponding discussion in \S 4.}.
\begin{figure}
   \includegraphics[width=\columnwidth]{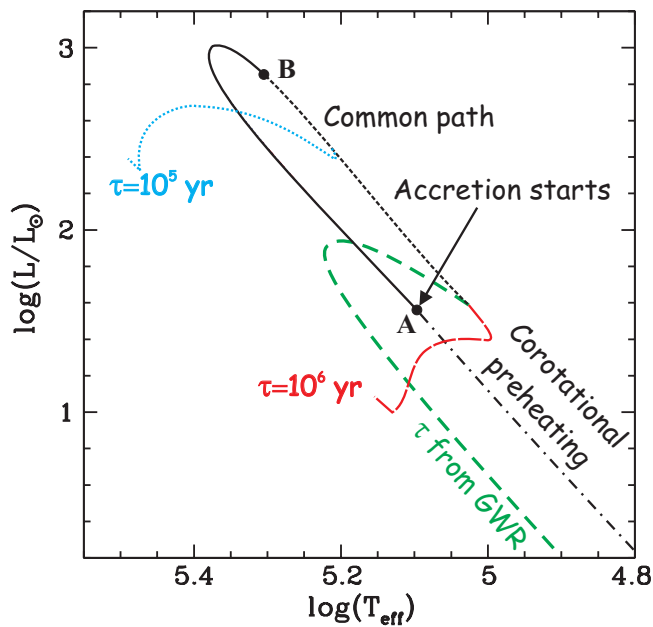}
     \caption{The evolution in the HR diagram of a 0.8 \msun\, rotating CO WD accreting mass from an accretion disk. The computation
                 has been halted when the WD attains the Chandrasekhar mass limit for rigidly rotating objects. Point {\bf A} marks the onset
                 of mass transfer, while point {\bf B} refers to the epoch when the critical angular velocity is attained on the surface of the
                 accretor. Dotted line refers to the model computed by assuming as braking timescale $\mathrm{\tau_B=10^{5}}$\,yr,
                 while the sequence depicted with a long dashed line have been obtained by fixing $\mathrm{\tau_B=10^{6}}$\,yr.
                 For the sake of comparison we plot also the model computed by \citet{piersanti2003b}, under the assumption that angular momentum is lost
                 via GWR after the accretor has deformed into a Jacobi ellipsoid (heavy dashed line).}
   \label{fig1}
\end{figure}

In Figure~\ref{fig1} we report the evolution in the Hertzsprung-Russell
(HR) diagram of the models computed with different braking timescales,
as labeled.  For comparison we report also the model computed by
assuming that angular momentum is lost from the accreting WD via
gravitational wave radiation emission already discussed in detail
in \citet{tornambe2013}.  In the figure, we display only the evolution
relative to the accretion phase, i.e. up to the point when the total
mass stored in the accretion disk has been transferred to the CO WD
and its total mass approaches the Chandrasekhar mass limit for rigidly
rotating degenerate objects.

Figure~\ref{fig1} clearly illustrates that the surface properties of
the accreting WD do depend on the angular momentum evolution, as
determined by both the mass deposition and the efficiency of the physical
mechanism decreasing the angular momentum. In particular,
independently of the assumed value for $\mathrm{\tau_B}$, after the
onset of the ``self-regulated'' accretion phase at point \textbf{A},
all the models follow the same evolutionary path, which is determined
only by the inward diffusion of the thermal energy excess produced on
the surface by mass deposition during the early phase of mass transfer
(from point \textbf{A} to \textbf{B} in Figure \ref{fig1}). 
During the ``self-regulated'' accretion phase the mass
transfer rate progressively decreases as the thermal balance along the
accreting WD is approached and the evolutionary timescale becomes
longer. The following evolution and, hence, the luminosity level at
which the accretor starts contracting and evolves blueward depend on
the assumed braking efficiency.  In particular, the higher the $\mathrm{\tau_B}$ 
value, the larger the braking efficiency, the higher
the corresponding value of the mass (and angular momentum) deposition
rate and, hence, the higher the luminosity level.  Figure~\ref{fig1}
also shows that for the model having braking efficiency determined
by the GWR emission the luminosity of the
accretor drops rapidly during the last part of the mass transfer
phase, when its total mass is approaching the Chandrasekhar mass limit
for rigidly rotating degenerate objects. In fact, according to the
discussion in \S 3.3 of \citet{piersanti2003b}, the GWR efficiency
drops rapidly, thus determining a rapid reduction of the effective
mass transfer rate. As a consequence, the thermal imbalance in the
accretor is completely smeared out by thermal diffusion active on a
timescale shorter than $10^7$ yr, so that the surface luminosity
rapidly decreases.  Model with $\tau_B=10^6$\,yr exhibits a moderate
decrease of the surface luminosity, while that with $\tau_B=10^5$\,yr
maintains almost the same luminosity level. According to the previous
considerations, in models with high braking efficiencies the thermal energy 
delivered in the surface layers during the early phase of mass transfer remains 
locally stored up to
the end of the mass deposition phase. This has an important consequence
for the expected observational properties during the following
evolutionary phase, when the WD attains the physical
conditions for carbon ignition in the inner zone where the degeneracy
level is very high (see Section 3).

It is worth emphasizing here that once the accretion process ceases, the total mass of 
the smaller WD has been accreted onto the larger WD leaving no circumstellar material 
to veil the surface of the remaining WD prior to its explosion.

We note that our results do not depend on the initial mass
of the accreting WD, as discussed in \citet[][see their
Figure~5]{tornambe2013}, because at each epoch the evolutionary
properties depend only on the assumed efficiency of angular momentum
losses, for a fixed value of the total mass of the accretor.

The results discussed above have been obtained assuming high
efficiency of angular momentum transport inside the accreting WD, so
that the accretor behaves as a rigid rotator. However, the same
considerations and results are still valid also when assuming 
that the angular momentum transport inside the CO WD is a secular
process, albeit with some relevant differences.  First of all, due to the
local storage of the deposited angular momentum in the external
layers, the accreting WD attains more rapidly the critical conditions;
moreover, as the spinning-up decreases the effective local gravity, the density in
the external layers is lower with respect to rigidly rotating
degenerate objects and, hence, the compressibility of the zone is
higher. As a consequence, when angular momentum transport is treated as
a secular process, the surface layers of the accretor heat up at a higher rate, so, as a
whole, the evolution in the HR diagram is quite similar to those
displayed in Figure~\ref{fig1} (from \textbf{A} to \textbf{B}), even if
the exact luminosity level as well as the corresponding effective
temperature at the epoch of the Roche instability (point \textbf{B} in
Figure~\ref{fig1}) depends on the efficiency of angular momentum
transport. During the ``self-regulated'' accretion phase, since a large
part of the deposited angular momentum remains localized in the
surface layers, the effective mass accretion rate decreases more
rapidly with respect to rigidly rotating objects; as a consequence,
the evolutionary timescale becomes longer and, hence, angular momentum
can be transferred inward efficiently. In fact, the
luminosity level at which the accretor starts evolving blueward, is
determined by the interplay between angular momentum losses and inward
angular momentum transport, which, in turn, determines the exact value
of the effective mass accretion rate. According to the previous
considerations, we can safely derive that, independently of the
angular momentum transport efficiency, the differential rotating WD
attains a lower luminosity for a fixed value of the efficiency of
angular momentum losses. Finally, we remark that, since the effective mass
transfer rate is lower and, hence, the evolutionary timescale is
longer, the thermal energy excess in the external layers is smeared out very rapidly,
as for the rigid rotating model losing angular momentum via
GWR. 
The occurrence of the luminosity drop, displayed in Figure~\ref{fig1} for 
the latter model, could not occur at all because the differential rotating WD
could never reach its own Chandrasekhar mass limit \footnote{We recall that, depending on the angular velocity
profile, the Chandrasekhar mass limit for a differentially rotating
degenerate object can exceed 5\,\msun.} during the mass and
angular momentum deposition phase.

\begin{figure}
   \includegraphics[width=\columnwidth]{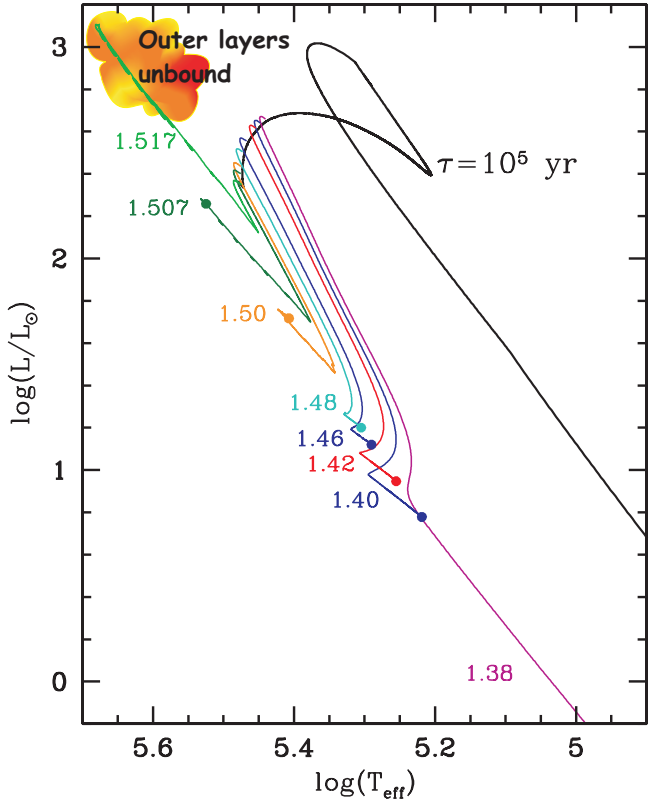}
   \caption{Evolution in the HR diagram of models with different total
     mass $\mathrm{M_F}$, as labeled, and the same efficiency of angular momentum
     losses. The evolution refers to the time spanning from the end of the
     mass transfer process and the epoch of the explosion, marked with
     an heavy dot.}
    \label{fig2}
\end{figure}

\section{The Final Path to the Explosion}

In the framework of the rotating DD scenario, all the available mass constituting the original donor and
forming the accretion disk around the accreting CO WD is completely
transferred without causing the C-ignition, and, hence, the explosion
as SN Ia. In fact, rotation decreases the effective gravity along the whole
structure so that a strong gravitational contraction as the total mass
increases is avoided. Therefore, when the mass reservoir has
been exhausted the resulting object is a stable massive WD whose
further evolution should be a slow and continuous cooling. However, it
has been suggested that these massive rotating objects are secularly
unstable and, hence, they should lose angular momentum via different
physical mechanisms \citep[see, e.g.][and references 
therein]{andersson1995,boshkayev2014,ilkov2012}. Owing to the reduction
of the total angular momentum, the WD undergoes a strong
compression so that it can heat up and ignite carbon in the innermost
zones in highly degenerate physical conditions.

To model this phase, once again we adopt an exponentially decaying law
for the total angular momentum and we parametrize the angular momentum
losses efficiency by using different timescale, as in the previous
section.  In Figure~\ref{fig2} we report the evolution in the HR diagram
of models computed by adopting $\mathrm{\tau_B=10^5}$\,yr and different
total masses of the initial binary system, as
labeled, from the end of the mass transfer process up to the explosion
epoch (marked with a dot).

\begin{figure}
\includegraphics[width=\columnwidth]{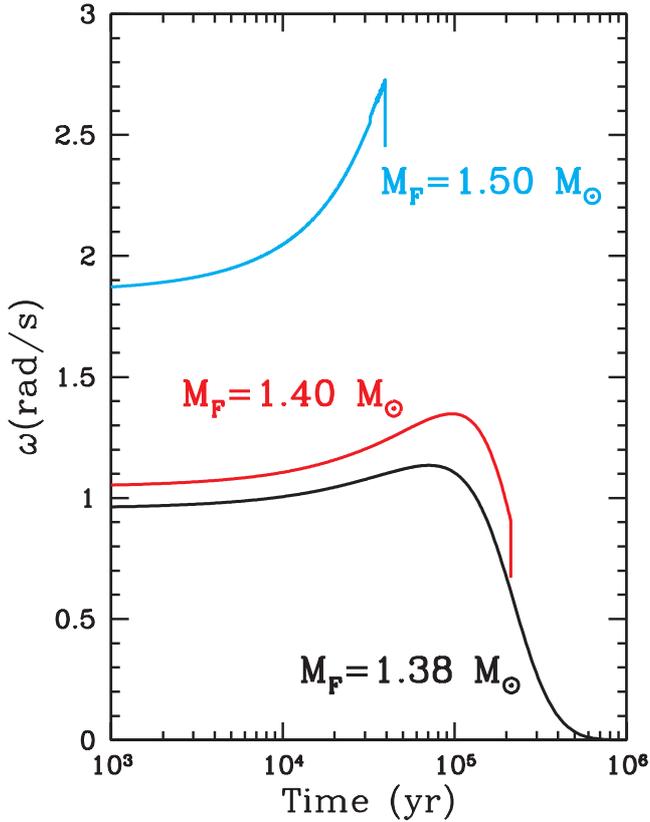}
\caption{Time evolution of the angular velocity $\omega$ for models
  with different total mass (as labeled), but the same efficiency of
  angular momentum losses. Time is set arbitrarily to zero at the epoch
  of the end of mass transfer process.}
   \label{fig3}
\end{figure}

It is worth noticing that, as already recognized
\citep{geroyannis2000,boshkayev2013}, when angular momentum is removed
from a rigidly rotating degenerate object more massive than the
canonical non-rotating Chandrasekhar mass limit, the structure reacts
by increasing its rotational velocity. This is clearly shown in
Figure~\ref{fig3}, where we report the time evolution of the angular
velocity for models losing angular momentum with
$\mathrm{\tau_B=10^5}$\,yr and different total mass, as labeled. Such a
behaviour can be easily understood when considering that rotating
stars are not solid but gaseous object; as a consequence, when angular
momentum is subtracted, the local effective gravity increases and the
star reacts contracting to achieve a new hydrostatic equilibrium
configuration and, hence, spins up. As a matter of fact, all the
models considered in Figure~\ref{fig2} experience an increase of the
angular velocity as angular momentum is lost. In any case the value of
the critical angular velocity on the surface increases more rapidly
than $\omega$ so that the structure is gravitationally stable.

As angular momentum continues to be subtracted, WDs with different
total mass experience different evolutionary paths. In particular, the
model with $\mathrm{M_{F}=1.38}$\,\msun\, becomes fully supported by
the pressure of degenerate electrons and, hence, cannot contract
further. Thus, the angular momentum losses remain the only physical
process driving the evolution of the angular velocity, which attains a
maximum and later on decreases continuously until the WD is
practically at rest.  Models more massive than 1.39\,\msun, but smaller
than 1.42\,\msun, experience the same evolution even if, owing to the
homologous compression of the whole structure, they succeed in
igniting carbon before the large part of the star attains highly
degenerate physical conditions (see model with
$\mathrm{M_{F}=1.40}$\,\msun\, in Figure~\ref{fig2}).  Finally, massive 
WDs contract so rapidly and, hence, heat up so efficiently, that
carbon is ignited when the star is still spinning up.  The drop in the
angular velocity value observed in the final part of the curves corresponding to 
models with total mass $\mathrm{M_{F}=1.40}$ and 1.50\,\msun\, 
after the accretion process is due to the
onset of central convection triggered by the huge energy released via
$\mathrm{{^{12}C}{+{^{12}C}}}$ reactions.
\begin{figure}
\includegraphics[width=\columnwidth]{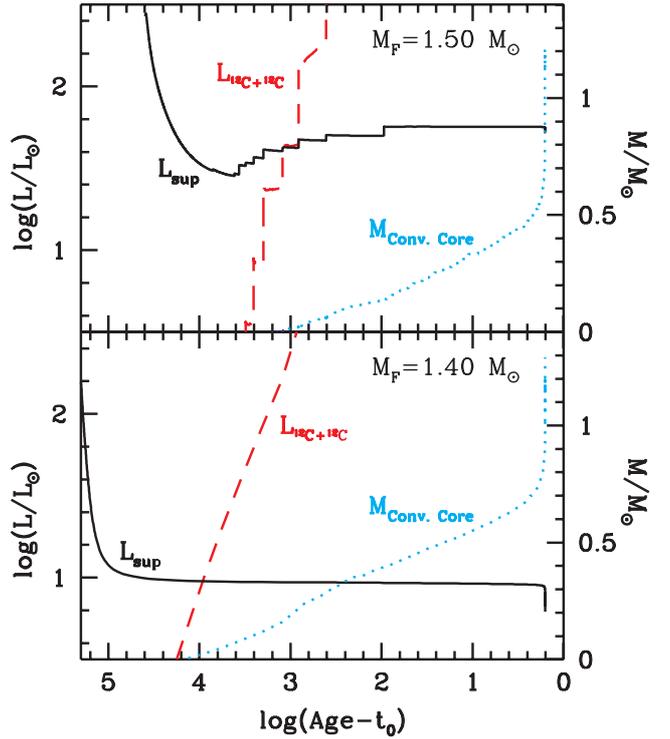}
\caption{Temporal evolution of the surface luminosity (solid line), of
  the carbon-burning luminosity (dashed line) and of the mass
  extension of the convective core (dotted line) for models with total
  mass $\mathrm{M_{F}=}$1.40 and 1.50\,\msun\, (lower and upper panel,
  respectively). Time (in yr) is expressed relative to the epoch of the
  explosion, which has been set to $\mathrm{t_0=2.1198\times 10^5}$\,yr
  and $\mathrm{3.9245\times 10^4}$\,yr for models in the lower and upper
  panel, respectively.  For both models we assume that angular
  momentum losses occur on the same timescale, namely $\tau_B=10^5$\,yr.}
   \label{fig4}
\end{figure}

The evolution of surface properties of the considered models after the
end of the accretion process and up to the explosion depends mainly on
the actual total mass of the accreted WD. In particular, after the
mass reservoir is exhausted and the evolution of the WD is driven by
the angular momentum loss, all the computed models experience a
luminosity drop (see Figure~\ref{fig2}). In this phase, owing to the
reduction of angular momentum, the accreted WD contracts, heating up
via homologous compression. In addition, as the evolutionary timescale
becomes longer, the thermal energy excess in the accreted layers
determined by the local compression via mass deposition, is
efficiently removed via inward thermal diffusion, while plasma
neutrino emission efficiently reduces the thermal content of the whole
star\footnote{Radiative losses are almost negligible in this phase}.
In models with total mass lower than 1.50\,\msun\,
the thermal diffusion occurs on a timescale shorter than the
compressional heating determined by the angular momentum losses, so
that the decrease of the surface luminosity continues up to when
$\mathrm{{^{12}C}+{^{12}C}}$ reactions are fully ignited
(corresponding to the point with the largest effective temperature in
Figure~\ref{fig2}). At variance, for larger masses, the luminosity starts to
increase before C-burning is ignited at a significant level. This 
different behaviour is due to the fact that more massive WDs are
closer to the rotating Chandrasekhar mass limit; as a consequence, the
contraction driven by angular momentum losses triggers a more rapid
and, hence, more efficient compressional heating of the whole star,
thus determining an increase of the surface luminosity. This occurrence
is illustrated in Figure~\ref{fig4} where we report results for 
models with total mass after accretion M=1.40 and M=1.50\,\msun\, (lower and upper panel,
respectively) the total surface luminosity ($\mathrm{L_{sup}}$, solid
lines), the luminosity produced by carbon fusion reactions
($\mathrm{L_{{^{12}C}+{^{12}C}}}$, dashed lines) and the mass
extension of the convective core ($\mathrm{M_{Conv. Core}}$, dotted
lines).  Figure~\ref{fig4} also shows that the energy (per unit of
time) produced via carbon burning at the center triggers very soon the
formation of a convective zone which rapidly grows in mass, involving
almost 95\% of the whole star. This determines a new change in the
physical properties of the star, which starts to expand due to the
fact that the thermal content in the whole convective region largely
deviates from that of a cold fully degenerate object. As a consequence
the star expands and, hence, as the evolutionary timescale becomes
shorter than the timescale on which angular momentum is removed, the
angular velocity starts to decrease (see the sudden drop in the
$\omega$ profiles for models with total mass $\mathrm{M_F}=$1.40 and 1.50\,\msun\,
displayed in Figure~\ref{fig3}. This corresponds to the last few days of
evolution before the explosion.

\begin{figure}
\includegraphics[width=\columnwidth]{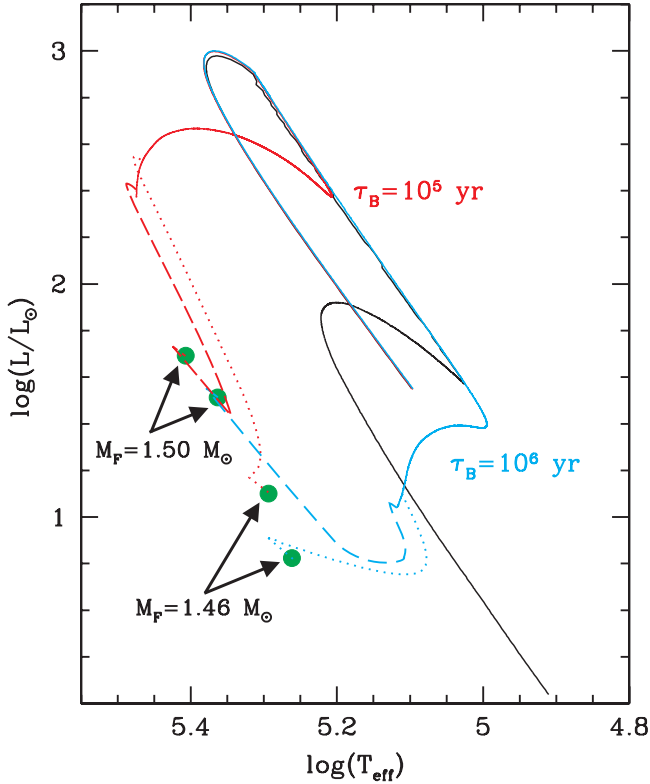}
\caption{Evolution in the HR diagram of models with the different
  total mass and efficiency of angular momentum losses.}
   \label{fig5}
\end{figure}

In Figure \ref{fig5} we compare
the evolution in the HR diagram of models with the same total mass,
namely 1.46\,\msun (dotted lines) and 1.50\,\msun (dashed lines) but with
different efficiency of angular momentum losses after the end of the
mass transfer phase. We note that, for a fixed value of the
total mass, the surface properties of the exploding objects are quite
similar, independently of the timescale adopted to remove angular
momentum. Such a finding is confirmed by several other computations
performed by varying both $\mathrm{M_F}$ and $\mathrm{\tau_B}$.

Finally, we note that during the last part of the
evolution, when the WD starts to expand due to the large energy
injected by C-burning, both $\omega$ and the critical angular velocity
on the surface decrease, but at a different rate, so that the star
remains gravitationally bound. The only model deviating from such a
behavior is the one with total mass $\mathrm{M_F=1.517}$\,\msun\,. In this
case the most external layers experience the Roche instability so that
it can be argued that $\simeq$ 0.01\,\msun\, of matter has to be lost
before the explosion could occur.

\section{About the Observational Properties}

In order to compare theoretical models with observational properties of SNe Ia 
progenitors, we assume genuine black bodies as well as spectral energy distribution 
(SED) models. To our knowledge, carbon and oxygen model atmospheres of such peculiar
WDs do not exist in the extant literature, so we adopt pure-Helium models from the
library available in the Tubingen non-local thermodynamic equilibrium (NLTE) Model-Atmosphere Package
(TMAP) database\footnote{http://astro.uni-tuebingen.de/$\sim$rauch/}
\citep{Rauch+03,vo:tmap_web}.

Table 1 lists the theoretical and the corresponding observable values for the set 
of rotating SNe Ia progenitors considered in the present work at the epoch of the 
explosion. We provide, for each model, the mass and luminosity (columns 1 and 2, solar units),
the effective temperature (column 3, in K), absolute AB magnitudes for
the Galex filters FUV and NUV (columns 4 and 5, respectively), the Sloan filters
(columns 6--10), two filters of the Advanced Camera for Surveys (ACS) on
board of the Hubble Space Telescope (HST), and, finally, the luminosity in two X-ray bands.

We perform the same evaluation by considering NLTE model atmospheres. Synthetic magnitudes 
in the GALEX, Sloan and ACS photometric bands, as well as for filters in the UVIS and IR channels 
of the Wide Field Camera 3 (WFC3) have been calculated using the iraf task {\it calcphot} of STSDAS 
and calibrated with WDs in the HST database\footnote{ftp://ftp.stsci.edu/cdbs/currentcalspec/}.
A selection of magnitudes is reported in Table~\ref{tab2}. As it can be easily noticed, 
pure He models are 0.3 -- 0.5 mag fainter than black-body models, the 
exact value depending on the mass 
of the exploding object and on the adopted braking efficiency.

\begin{table*}
  \caption{Expected AB absolute magnitudes in various bands for SNe Ia progenitors just before explosion based on black-body models. The different mass of the considered
    models are determined by the rotationally-driven accretion mechanism as suggested by \citet{piersanti2003b}. Lines 1--7 refer to models 
    computed by adopting as braking timescale $10^5$ yr and displayed in Figure \ref{fig2}, while lines 8--9 are relative to models computed with braking timescale $10^6$ yr 
    and displayed in Figure \ref{fig5}.
}
	\label{tab1}
    \centering
    \begin{tabular}{ cc c c c c c c c c c c c c }
      \hline\hline
      Mass (${\rm M_\odot}$) & $\mathrm{\log(\frac{L}{L_\odot})}$ & $\mathrm{\log(T_{eff})}$ &  FUV & NUV & $u$ & $g$ & $r$ & $i$ & $z$ & F435W & F814W & 0.5-2\,keV & 2-10\,keV\\
      \hline
      \multicolumn{14}{c}{$\mathrm \tau_{B}=10^5$ yr} \\
      \hline
      1.400 & 0.77 & 5.22 & 8.14 & 8.87 & 9.78 & 10.33 & 10.92 & 11.32 & 11.70 & 10.18 & 11.48 & 1.67E+23 & 6.34E-25\\
      1.420 & 0.94 & 5.26 & 7.98 & 8.73 & 9.64 & 10.20 & 10.78 & 11.19 & 11.57 & 10.04 & 11.34 & 3.91E+24 & 3.59E-19\\
      1.460 & 1.11 & 5.29 & 7.76 & 8.51 & 9.43 & 10.00 & 10.58 & 10.99 & 11.36 &  9.83 & 11.14 & 3.87E+25 & 3.79E-15\\
      1.480 & 1.19 & 5.31 & 7.70 & 8.45 & 9.38 &  9.94 & 10.53 & 10.94 & 11.31 &  9.78 & 11.09 & 1.52E+26 & 1.20E-12\\
      1.490 & 1.26 & 5.32 & 7.59 & 8.35 & 9.27 &  9.84 & 10.42 & 10.84 & 11.21 &  9.68 & 10.99 & 3.17E+26 & 2.07E-11\\
      1.500 & 1.71 & 5.41 & 7.09 & 7.87 & 8.80 &  9.37 &  9.96 & 10.38 & 10.75 &  9.21 & 10.53 & 8.28E+28 & 1.29E-01\\
      1.507 & 2.25 & 5.52 & 6.52 & 7.31 & 8.26 &  8.83 &  9.43 &  9.84 & 10.22 &  8.67 & 10.00 & 2.03E+31 & 4.37E+08\\
      \hline
\multicolumn{14}{c}{$\mathrm \tau_{B}=10^6$ yr} \\
\hline
 1.460 & 0.83 & 5.26 & 8.26 & 9.00 & 9.91 & 10.47 & 11.06 & 11.47 & 11.84 & 10.32 & 11.62 & 3.04E+24 & 2.79E-19\\
 1.500 & 1.53 & 5.36 & 7.19 & 7.96 & 8.89 &  9.45 & 10.04 & 10.50 & 10.83 &  9.30 & 10.60 & 5.04E+27 & 9.73E-07\\
 \hline
\end{tabular}
\end{table*}

\begin{table*}
  \caption{AB magnitudes for selected filters as derived from NLTE atmospheres models.}

	\label{tab2}
    \centering
    \footnotesize
    \begin{tabular}{ ccccccccccccccc }
      \hline\hline
      Mass (${\rm M_\odot}$) &      V    & FUV  & NUV  &  $u$   & $g$    & $r$  & $i$   & $z$   & F225W & F336W & F475W & F555W & F814W & F110W \\
      \hline
      \multicolumn{14}{c}{$\mathrm \tau_{B}=10^5$ yr} \\
      \hline
 1.400 &   11.00 &  8.40 & 9.17 & 10.10  & 10.66 & 11.25 & 11.65 & 12.03 & 9.26 & 9.98  & 10.69  & 10.90 & 11.78 & 12.51 \\
 1.420 &   10.87 &  8.27 & 9.05 & 10.00  & 10.54 & 11.12 & 11.53 & 11.90 & 9.14 & 9.86  & 10.57  & 10.78 & 11.66 & 12.38 \\
 1.460 &   10.68 &  8.08 & 8.85 &  9.78  & 10.35 & 10.93 & 11.34 & 11.71 & 8.95 & 9.67  & 10.38  & 10.59 & 11.46 & 12.18 \\
 1.480 &   10.64 &  8.04 & 8.81 &  9.74  & 10.31 & 10.89 & 11.30 & 11.67 & 8.91 & 9.63  & 10.34  & 10.55 & 11.42 & 12.14 \\
 1.490 &   10.55 &  7.94 & 8.72 &  9.65  & 10.21 & 10.80 & 11.20 & 11.57 & 8.81 & 9.54  & 10.24  & 10.46 & 11.33 & 12.05 \\
 1.500 &   10.14 &  7.54 & 8.32 &  9.25  &  9.81 & 10.39 & 10.80 & 11.16 & 8.41 & 9.13  &  9.84  & 10.05 & 10.92 & 11.64 \\
 1.507 &   9.65  &  7.04 & 7.82 &  8.76  &  9.32 & 9.90  & 10.30 & 10.67 & 7.91 & 8.64  &  9.35  &  9.56 & 10.43 & 11.14 \\ 
\hline
\multicolumn{14}{c}{$\mathrm \tau_{B}=10^6$ yr} \\
 1.460 &   11.15 &  8.55 & 9.33 & 10.25  & 10.82 & 11.40 & 11.81 & 12.18 & 9.42 & 10.14 & 10.85  & 11.06 & 11.94 & 12.66 \\ 
 1.500 &   10.19 &  7.59 & 8.37 &  9.30  &  9.86 & 10.44 & 10.85 & 11.22 & 8.46 & 9.18  &  9.89  & 10.10 & 10.98 & 11.70 \\
\hline
\end{tabular}
\end{table*}

The values listed in Table \ref{tab1} and \ref{tab2} span the whole range of expected magnitudes for SNe Ia progenitors 
close to the explosion in the framework of the rotationally-driven scenario discussed in the previous 
sections. We emphasize again that the results are independent of the assumed scenario for the 
progenitors (SD, DD, CD scenarios) but they are determined mainly by the total mass of the exploding object 
and, to a lesser extent, on the physical mechanism driving the braking and/or the angular momentum redistribution 
during the accretion phase. 
Inspection of these two tables reveals that the spectral distribution is peaked in the far-UV/soft-X range. 
Moreover, they also show that the related optical emission would be so weak that only nearby events would be 
detectable.

\section{Discussion and final remarks}

The recent observations of SN 2011fe in M~101 \citep{Nugent+11a,Nugent+11b} 
and SN~2014J in M~82 \citep{fossey2014,zheng+2014} have stimulated a large debate 
about the possibility to constrain observationally their progenitor systems. 
 
Several studies have reported observational features and suggested 
possible progenitor systems \citep[e.g.][]{Li+11,Kelly+14,McCully+14}. 
Observations at different wavelengths have suggested that some progenitor 
systems associated with the SD scenario have to be ruled out for
SN~2011fe, leaving a very limited phase space compatible with the DD scenario 
and really exotic single degenerate systems \citep{Li+11,Chomiuk+12}.
Optical nebular spectra do not show any evidence for narrow
H$\alpha$ emission, supporting again a degenerate companion for
SN~2011fe. A pre-explosion frame in the HST archive exhibits the same 
properties, so that red giants and most helium-star companions can be safely 
ruled out \citep{Li+11}. 
By using pre-explosive archival HST images of M~82 from the near-UV to the near-IR, 
\citet{Kelly+14}, excluded for SN~2014J a progenitor system having a bright red giant donor,
including recurrent novae with luminosities comparable to the Galactic
prototype symbiotic system RS Oph. According to the available data, 
these authors also remark that a system consisting of two CO WDs as well as a system 
made by a CO WD and a main sequence star can not be excluded.

In the framework of the rotationally-driven scenario described in the previous sections, it is not surprising 
that even in the case of nearby SN 2011fe and SN 2014J the imprint of the progenitor system can not be detected in 
archival frames. In this regard we note that the SN2011fe progenitor 
should have a magnitude in ACS/F435W band $\sim$38 mag or higher, while the current observational limit 
is 27.4\, mag \citep{Li+11}. Also the largest possible X-ray emission in the 0.5-2
keV band, corresponding to the model with mass 1.507\msun\, (see Table \ref{tab1}), is several orders of
magnitude lower than the upper limit inferred for the SN 2014J by
\citet{nielsen2014}. In fact, only SNe Ia progenitors in our own Galaxy
may be detectable.

\begin{figure}
   \includegraphics[width=\columnwidth]{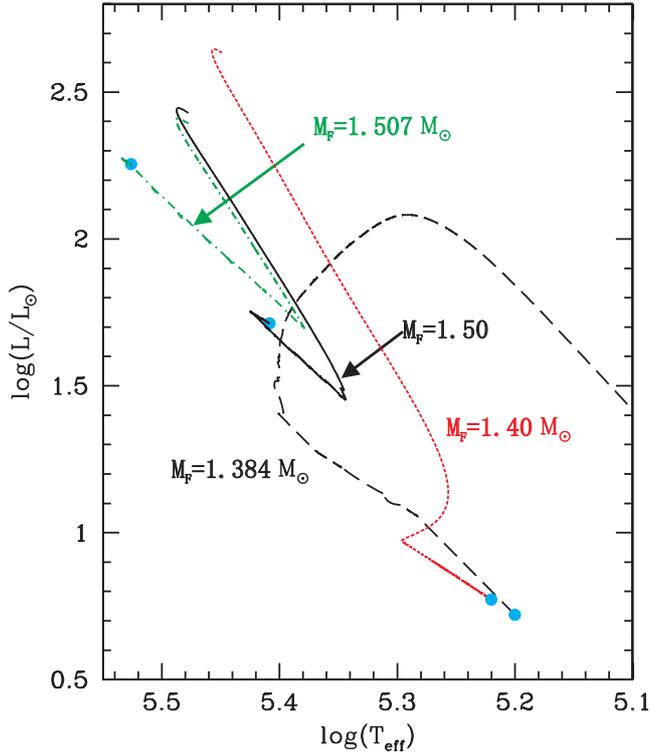}
  \caption{Comparison between a classical non-rotating standard model arising in a Single Degenerate 
    binary system (long dashed line - $M_F=1.384$\msun) and selected models reported in Figure \ref{fig2}, 
    having $\tau_{B}=10^5$ yr and different masses, as labeled.}
    \label{fig6}
\end{figure}

As a final consideration, we note that, according to predictions in Table \ref{tab1} and \ref{tab2}, models 
accounting for the effects of rotation exhibit a strong feature in the 
0.5-2 keV band, having an expected flux which depends on the total mass of the progenitor. 
Such a feature is completely missing in standard non-rotating Chandrasekhar mass models of SNe Ia 
owing to their lower effective temperature. Such an occurrence is clearly illustrated in Figure \ref{fig6}, 
where we compare models having $\tau_{B}=10^5$ and different total masses with the ZSUN 
model in \citet{piersanti2017}, obtained by accreting directly CO-rich matter onto an initial 0.817\msun\, 
CO WD at \mdot=$10^{-7}$\msun\pyr, mimicking the evolution of a SD system. 
For the latter object, the expected flux in the 0.5--2 keV band mass is
vanishingly small whereas it becomes appreciable for the more massive models
obtained in the framework of the rotationally-driven scenario. As a
consequence it seems that the X ray flux, even if small, can be used to assess 
the role played by rotation in the evolution of SNe Ia progenitors.

\section*{Acknowledgements}

It is a pleasure to thank several colleagues for their advices 
comments, such as I. Dom{\'i}nguez, C. Badenes, B. Fisher, E. Cappellaro, M. Turatto,  
J. Isern, S. Benetti, E. Bravo, F. Mannucci, M. Dall'Ora, M. Della Valle, E. Garcia-Berro. 
We want also to thank the referee, J. Danziger, for his useful comments and suggestions 
which improve the presentation of our results.
This work received partial
financial support by INAF$-$PRIN$/$2014 (PI G. Clementini). The
TheoSSA service (http://dc.g-vo.org/theossa) used to retrieve
theoretical spectra in this work was constructed as part of the
activities of the German Astrophysical Virtual Observatory.



\bibliographystyle{mnras}
\bibliography{tornambe} 


%

\bsp	
\label{lastpage}
\end{document}